\documentclass[a4paper]{article}

\usepackage[english]{babel}
\usepackage[utf8x]{inputenc}
\usepackage[T1]{fontenc}
\usepackage{array}

\usepackage[a4paper,top=3cm,bottom=2cm,left=3cm,right=3cm,marginparwidth=1.75cm]{geometry}

\usepackage{amsmath}
\usepackage{graphicx}
\usepackage[colorinlistoftodos]{todonotes}
\usepackage{cite}
\usepackage[colorlinks=true, allcolors=blue]{hyperref}
\usepackage{authblk}

\begin{document}
\title{Naturally occurring fluorescence in transparent insect wings}

\author[a,b,*]{S\'{e}bastien R. Mouchet}
\author[c,*]{Charlotte Verstraete}
\author[d]{Bojana Bokic}
\author[c,e]{Dimitrije Mara}
\author[b]{Louis Dellieu}
\author[f]{Albert G. Orr}
\author[b]{Olivier Deparis}
\author[e]{Rik Van Deun}
\author[c]{Thierry Verbiest}
\author[a]{Pete Vukusic}
\author[d,g]{Branko Kolaric}
\affil[a]{School of Physics, University of Exeter, Stocker Road, Exeter EX4 4QL, United Kingdom}
\affil[b]{Department of Physics \& Namur Institute of Structured Matter (NISM), University of Namur, Rue de Bruxelles 61, 5000 Namur, Belgium}
\affil[c]{Molecular Imaging and Photonics, Department of Chemistry, KU Leuven, Celestijnenlaan 200D, 3001 Heverlee, Belgium}
\affil[d]{Center for Photonics, Institute of Physics, University of Belgrade, Pregrevica 118, 11080 Belgrade, Serbia}
\affil[e]{L$^{3}$ \textendash~Luminescent Lanthanide Lab, Department of Chemistry, Ghent University, Krijgslaan 281-S3, 9000 Ghent, Belgium}
\affil[f]{Environmental Futures Centre, Griffith University, Nathan, QLD 4111, Australia}
\affil[g]{Micro- and Nanophotonic Materials Group, University of Mons, Place du Parc 20, 7000 Mons, Belgium}
\affil[*]{Co-shared first authorship}
\maketitle

\section{Introduction}

\begin{figure}
	\includegraphics[width=\columnwidth]{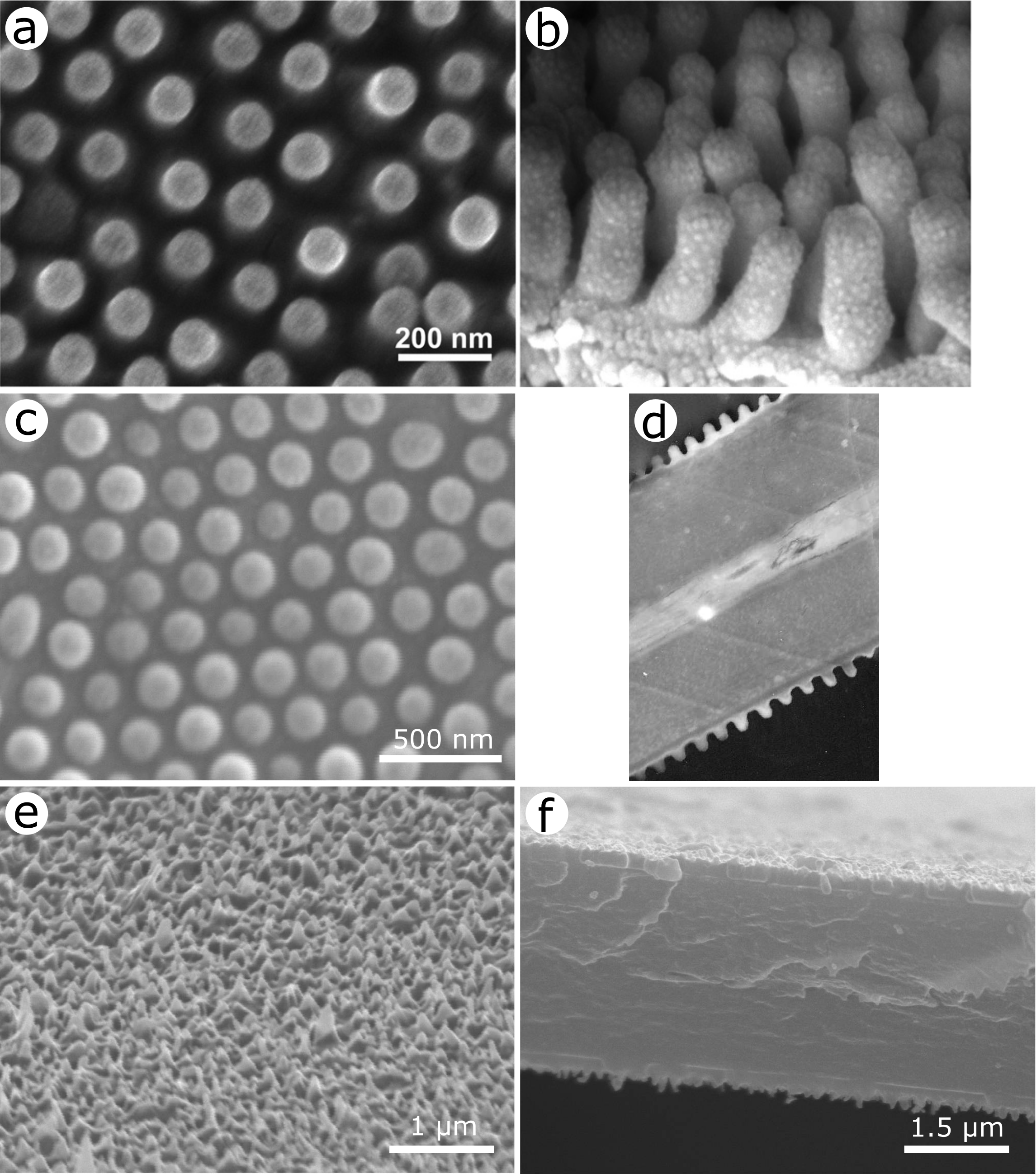}
	\caption{\textbf{The transparent parts of the wings of insects including the grey cicada \textit{Cicada orni} (a,b), the broad-bordered bee hawk-moth \textit{Hemaris fuciformis} (c,d) and the dragonfly blue hawker \textit{Aeshna cyanea} (e,f) are covered by arrays of protrusions.} These protrusions are sometimes displayed along a quasi-periodic hexagonally close-packed array such as on the wings of \textit{C. orni} (a,b) and \textit{H. fuciformis} (c) or a disordered pattern such as in the case of \textit{A. cyanea} (e), as observed here by scanning electron microscopy (a-c,e-f) and transmission electron microscopy (d) with top views (a,c), oblique views (b,e) and views of cross-sections (d,f). Figures (a,b) and (d) were reproduced (a,b) from ref.~\cite{Dellieu2014} with permission from AIP publishing and (d) from ref.~\cite{Vukusic2003} with permission from Springer Nature.
	}
	\label{fig:SEM-TEM}
\end{figure}

Fluorescence emission takes place in the integuments of several natural organisms including species from the animal classes Insecta, Aves, Amphibia, Reptilia and Mammalia as well as the plant kingdom~\cite{Pavan1954,Tani2004,Vukusic2005,Iriel2010a,Iriel2010b,Welch2012,Lagorio2015,Gruber2015,Gruber2016,Mouchet2016a,Marshall2017,Taboada2017,Deschepper2018,Mouchet2019,Ladouce2020,MohdTop2020,Croce2021,Reinhold2021,Toussaint2021}. This emission of light has been shown to play a role in the visual communication of some organisms, facilitating species recognition, mate selection, prey detection, camouflage and agonistic behaviour~\cite{Lagorio2015,Marshall2017}. Fluorescence results from the presence of fluorophores, such as papiliochrome II, biopterin, psittacofulvin, green fluorescent protein (GFP) within the biological integuments. Following the absorption of incident light (typically in the UV or the shorter-wavelength part of the visible range), these molecules emit light at a longer wavelength (usually in the visible), resulting from transitions between their electron states. In spite of the crucial role it is believed to play in nature, fluorescence in natural organisms remains under-investigated from optical, chemical and biological perspectives. One example is the transparent wings of insects from the order Hemiptera. For example, in both males and females of many cicadas, both pairs of wings are known for their anti-reflective properties~\cite{Stoddart2006,Sun2011,Dellieu2014,Deparis2014,Verstraete2018linear} arising from electromagnetic impedance matching between the air and the wing material covered by quasi-periodic arrays of hexagonally close-packed protrusions (Fig.~\ref{fig:SEM-TEM}a,b). Similar structures also cover parts of the integument in other organisms such as the wings of several moths and butterflies (Order Lepidoptera), dragonflies and damselflies (Order Odonata) (Fig.~\ref{fig:SEM-TEM}c-f)~\cite{Yoshida1996,Yoshida1997,Yoshida2002,Vukusic2003,Hooper2006,Deparis2009,Deparis2014,Stavenga2014,Siddique2015,Mouchet2018} and the corneas of some insect eyes~\cite{Bernhard1965,Stavenga2006}. In all these examples, the basic building material of the insect tissues is chitin, a polysaccharide, overlaid by a thin epicuticle of wax. In cicadas, transmittance spectra of their wings have been measured to reach more than 90\% and to peak at up to 98\% over the range from 500 to 2500~nm~\cite{Stoddart2006,Sun2011}. These anti-reflective properties are thought to be involved as part of the insects’ cryptic strategy and the nanostructured protrusions have been mimicked for the development of anti-reflective coatings for a full range of applications such as solar cells, screens, anti-glare glasses, light-sensitive detectors, telescopes and camera lenses~\cite{Xie2008,Han2016,Zada2017}. In addition to this optical property, the transparent wings of insects are known to be hydrophobic~\cite{Zhang2006,Sun2009,Sun2011,Wisdom2013,Dellieu2014,Deparis2014,RomanKustas2020} and some exhibit anti-bacterial behaviour~\cite{RomanKustas2020,Ivanova2012,Ivanova2013,Diu2014,Kelleher2015}. In the case of cicadas, the surface morphology was found to influence jointly the hydrophobic and anti-reflective properties (Fig.~\ref{fig:SEM-TEM})~\cite{Dellieu2014,Deparis2014}. The protrusions on the wing surface can be modelled as truncated cones covered by hemispheres. The cone shape affects the transparency and the hemispheres enhance the wing's hydrophobicity.
Despite these properties, which are relevant in many scientific fields ranging from physics and material science to chemistry and biology, fluorescence emission has so far not been reported in any of the 3,000 described species of the superfamily Cicadoidea. The phenomenon has however been investigated in several butterfly and moth species, in which the coloured wing scales are known to embed fluorophores such as papiliochrome II~\cite{Cockayne1924,Kumazawa1994,Vukusic2005,Trzeciak2012,Welch2012,Wilts2012}. Photonic structures present in these scales may mediate the associated light emission~\cite{Vigneron2008,VanHooijdonk2011,VanHooijdonk2012c}. Fluorescence has also been described in the vein joints and the membrane of the transparent wings of damselflies and dragonflies~\cite{Gorb1999,Appel2011,Appel2015,Chuang2016}. This emission was attributed to the presence of autofluorescent proteins such as resilin. Resilin proteins emit blue light (at about 400~nm) under incident UV light and were reported to decrease in concentration toward the distal regions of the wings (on both sides)~\cite{Gorb1999,Appel2011}. They give rise to low stiffness and high strain in the biological tissues, allowing passive wing deformations~\cite{Elvin2005,Burrows2008}. We note that resilin fluorescence may often be incidental with little or no semiotic function. However, in the case of males of the red-winged damselfly \textit{Mnesarete pudica} (Hagen in Selys, 1853), fluorescence emission from resilin-containing wings was demonstrated to have a role in signalling (e.g., in courtship and territorial behaviour) in combination with pigments and UV reflection, by modulating the visual appearance of the insect according to sex and age~\cite{Guillermo-Ferreira2014}.

\begin{figure}
	\includegraphics[width=\columnwidth]{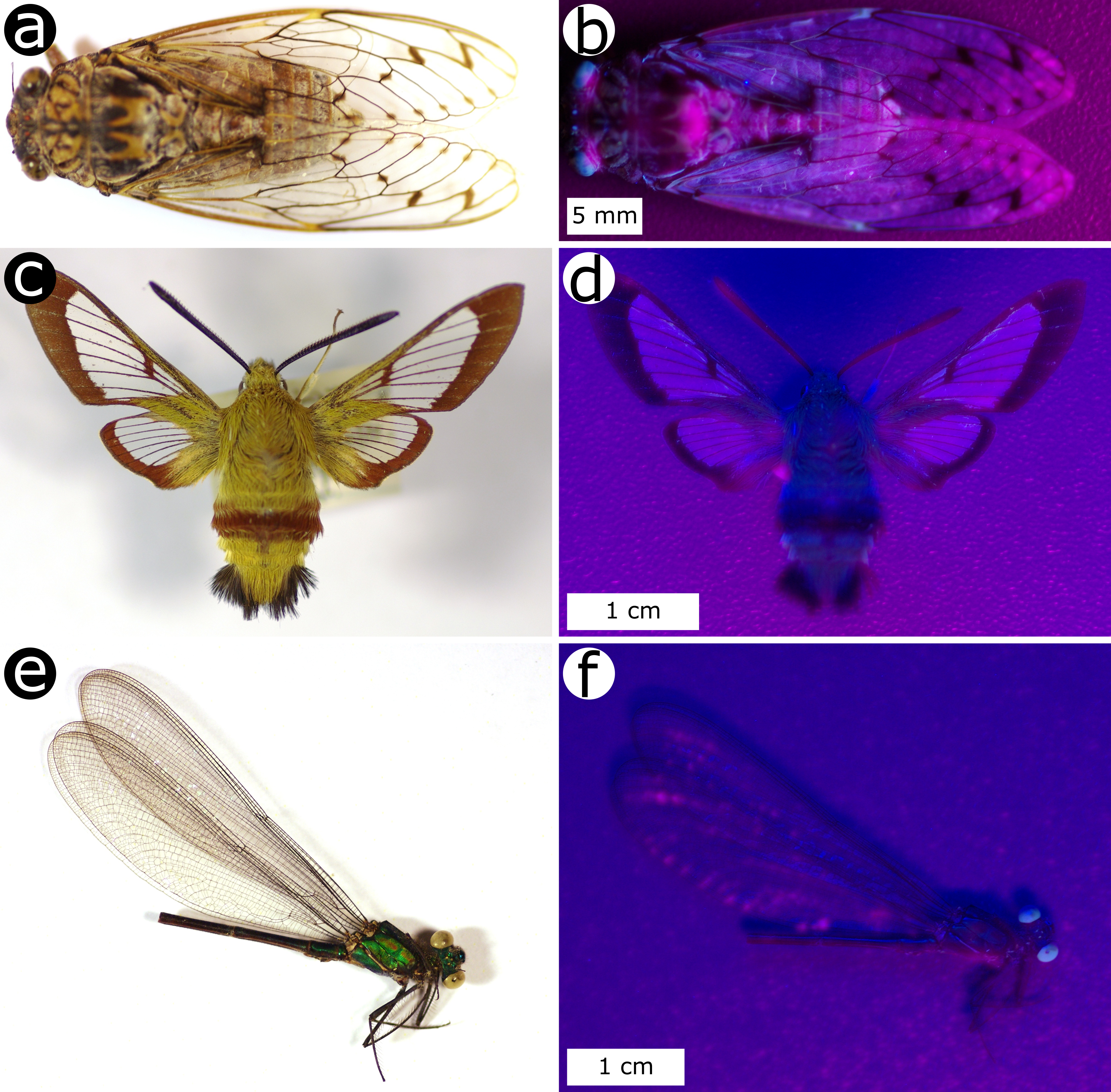}
	\caption{\textbf{The wings of the grey cicada \textit{C. orni} (a), the broad-bordered bee hawk-moth \textit{H. fuciformis} (c,d) and the damselfly \textit{V. amabilis} (e,f) emit light by fluorescence under UV irradiation.} \textit{C. orni} has a brown-grey body and transparent wings under visible incident light (a). The latter display a bluish hue under UV light (b). Similarly, \textit{H. fuciformis} and \textit{V. amabilis} transparent wings (c,e) exhibit a blue colour due to fluorescence emission (d,f).}
	\label{fig:picture_C_orni_H_fuciformis_V_amabilis}
\end{figure}

In this article, we characterised the fluorescence emission from the transparent wings of two species of cicadas, the grey cicada \textit{Cicada orni} (Linnaeus, 1758) (Fig.~\ref{fig:picture_C_orni_H_fuciformis_V_amabilis}a,b) and the common cicada \textit{Lyristes (Tibicen) plebejus} (Scopoli, 1763), using both linear and nonlinear optical methods. We compared their properties with the fluorescence emission from two other insects exhibiting transparent wings, namely, \textit{Hemaris fuciformis} (Linnaeus, 1758) commonly known as the broad-bordered bee hawk-moth (Fig.~\ref{fig:picture_C_orni_H_fuciformis_V_amabilis}c,d) and the Bornean damselfly \textit{Vestalis amabilis} Lieftinck, 1965 (Fig.~\ref{fig:picture_C_orni_H_fuciformis_V_amabilis}e,f). These fluorescent behaviours were optically characterised by one- and multi-photon microscopy and spectrometry, allowing us to infer the symmetry of the material as well as the biological role of the fluorophores in the visual perception of insects and to explain the previously measured high absorption in the range 300-400~nm~\cite{Dellieu2014}.

\section{Materials and Methods}

\subsection{Sample collection}
Adult specimens of \textit{C. orni} and \textit{L. plebejus} were collected in Fournès in the Gard Department (France) on the 22nd of June 2014 (by S.R. Mouchet and L. Dellieu), and \textit{V. amabilis} was collected in swamp forest in Belait Division (Brunei Darussalam) in 1994 (by A.G. Orr). Adult specimens of \textit{H. fuciformis} were bought from licensed vendors. \textit{V. amabilis} was immersed for 24 hours in with acetone and air-dried. Other specimens were air-dried. No further sample preparation was necessary for the microscopy and spectrofluorimetry observations. All analyses were performed on the insects' wings.

\subsection{Optical and fluorescence microscopy}
The sample wings were analysed using two different configurations of the microscope: a reflection mode (illumination with visible light and observation in the visible range) and a fluorescence mode (illumination with ultraviolet light and observation in the visible range).
Optical microscopy analysis was performed using an Olympus BX61 (Tokyo, Japan) microscope, an Olympus XC50 camera and an Olympus BX-UCB visible light source (in reflection mode) or a Lumen Dynamics X-cite Series 120PCQ (Mississauga, Ontario, Canada) UV-lamp (in fluorescence mode).

\subsection{Spectrofluorimetry}
One photon fluorescence (1PF) emission spectra were recorded using a 450~W xenon lamp as steady state excitation source. The incident light formed a 45° angle with the normal direction to the sample surface and emitted light was detected at a 45° angle on the other side of the normal direction. The sample was excited at different excitation wavelengths: 320~nm, 340~nm, 360~nm, 380~nm, 400~nm, 420~nm and 440~nm. Time-resolved measurements of fluoresence emission were performed using a microF920H xenon Flashlamp light source operating at a frequency of 100~Hz with a pulse width of 4~µs and excitation wavelengths equal to 340~nm and 360~nm. These time-resolved measurements were performed at the wavelengths of the maximum of emission peak intensity. They were fitted to double exponential functions. These best fits allowed us to assess the decay time of the fluorescence emission.

\subsection{Multiphoton microscopy}
Multiphoton microscopy experiments were carried out with an Olympus (M\"{u}nster, Germany) BX61 WI-FV1200-M system. The laser was a Spectra-Physics (Santa Clara, CA, USA) InSight DS+ laser (82-MHz repetition rate, 120-fs pulse width, p-polarised) at 800~nm, 900~nm, 1000~nm and 1100~nm fundamental wavelengths. The incident laser power was controlled by an achromatic half-wave plate and a s-oriented polariser directly after the laser. The laser beam was focused on the sample via either a 15X LMV objective (NA 0.3) or a 50X SLMPlan N objective (NA 0.35). The resulting signal was detected non-descanned in backwards reflection via a Hamamatsu R3896 photomultiplier tube. Different detection optics were used, depending on the excitation wavelength, in order to separate the two-photon fluorescence (2PF) and the Second Harmonic Generation (SHG) signal. At 1000~nm, a 525~LPXR dichroic mirror was used to remove the SHG from the 2PF, and an additional 500/10 bandpass filter cleared the SHG signal further. Similar filters were used for the other three fundamental wavelengths. The signal at each pixel was depicted in the micrographs as different intensities in false green (SHG) or false red (2PF) colours. A routine providing an extended depth field allowed us to obtain in-focus micrographs from multiple images recorded at different depths in steps of 4~µm. Samples were observed over several minutes of time, in order to assess the photostability.

\subsection{Electron microscopy}

Apart from two scanning electron microscopy (SEM) images taken from the literature~\cite{Dellieu2014}, SEM observations were performed with a FEI (Hillsboro, OR) Nova 600 NanoLab Dual-Beam FIB/SEM microscope on small transparent regions of the wing membranes of \textit{H. fuciformis} and \textit{A. cyanea}. These samples were mounted onto SEM stubs with electrically conducting epoxy resin and sputter-coated with ca. 8 nm of gold palladium.

Transmission electron microscopy (TEM) imaging was carried on with cross-sections of transparent regions of \textit{H. fuciformis}'s wings using a JEOL (Tokyo, Japan) 100S TEM instrument. The samples were fixed in 3\% glutaraldehyde at 21\textdegree C for 2 hours and consequently rinsed in sodium cacodylate buffer before being fixed in 1\% osmic acid in buffer for 1 hour. The samples were then block-stained in 2\% aqueous uranyl acetate for 1 hour, dehydrated through an acetone series (ending in 100\% acetone) and embedded in Spurr resin~\cite{Spurr1969}. After microtoming, cross-sections were stained with lead citrate.

\section{Results and Discussion}

\begin{figure}
    \centering
	\includegraphics[width=0.8\columnwidth]{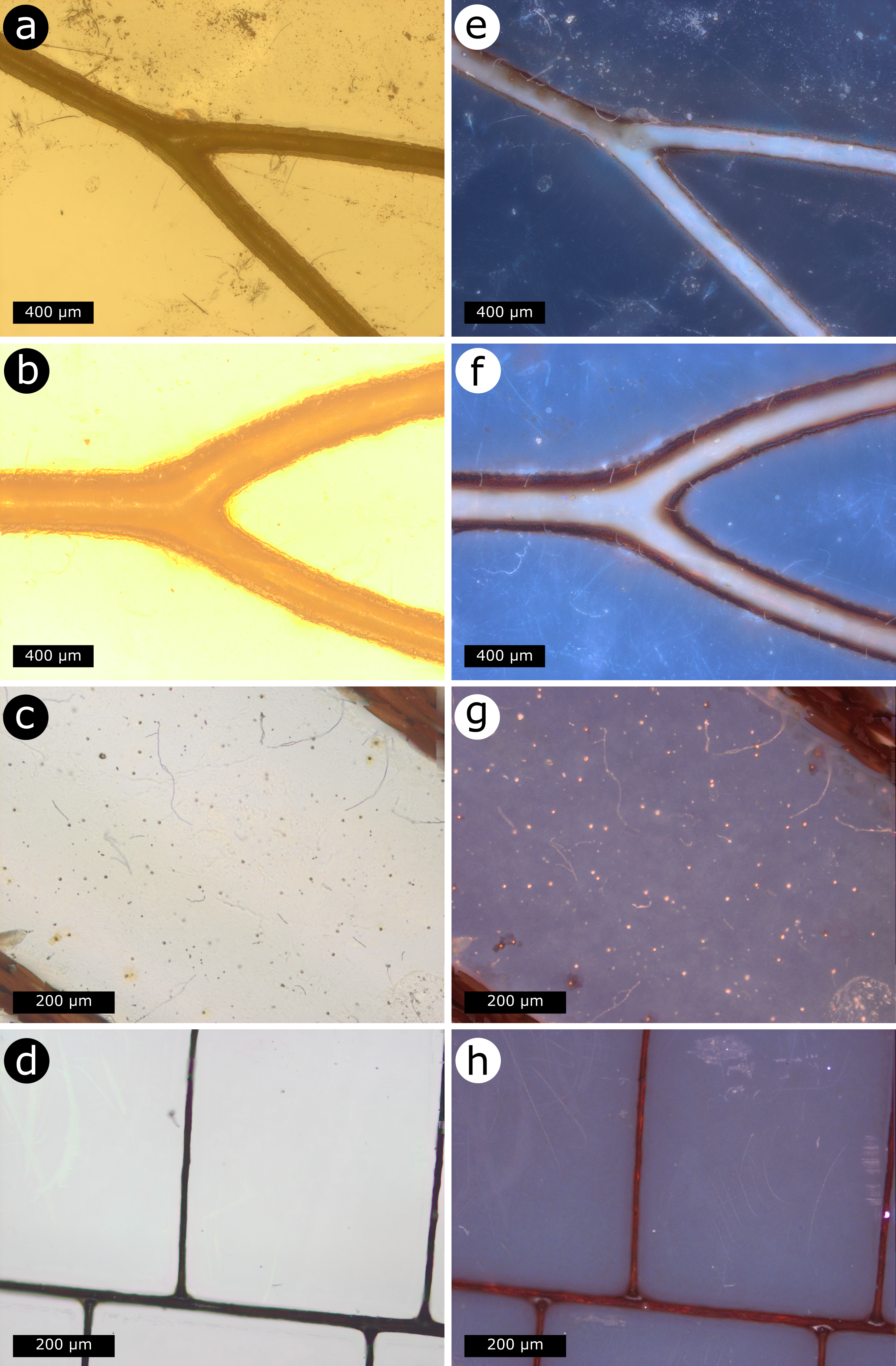}
	\caption{\textbf{The membranes and the veins of the transparent wings of the grey cicada (\textit{Cicada orni} (Linnaeus, 1758)), the common cicada (\textit{Lyristes (Tibicen) plebejus} (Scopoli, 1763)), the broad-bordered bee hawk-moth (\textit{Hemaris fuciformis} (Linnaeus, 1758)) and \textit{Vestalis amabilis} damselfly (Linnaeus, 1758) emit light by fluorescence under UV irradiation.} The transparent wings of \textit{C. orni} (a), \textit{L. plebejus} (b), \textit{H. fuciformis} (c) and \textit{V. amabilis} (d) are divided by brown veins. Both their transparent membranes and the veins of the wings emit visible light under UV incident light (e-h). The 1PF signal from the veins of the cicada species (e,f) is specifically intense.}
	\label{fig:Micrograph_1PF}
\end{figure}

Both \textit{C. orni} and \textit{L. plebejus} cicada species have brown-grey bodies and transparent wings (Fig.~\ref{fig:picture_C_orni_H_fuciformis_V_amabilis}) with very low light reflection~\cite{Stoddart2006,Sun2011,Dellieu2014,Deparis2014,Verstraete2018linear}. The bodies and wings are approximately 25~mm and 30-35~mm long, respectively. Like most insects including moths and damselflies, the wings of cicadas are divided by veins (Fig.~\ref{fig:Micrograph_1PF}a-d) appearing as darker structures forming the structurally flexible frame supporting of the wing membranes, while the transparent parts are membranes that form the main aerodynamic surface of the wing~\cite{Elvin2005,gullion2017}. 


Upon illumination with UV light (Fig.~\ref{fig:picture_C_orni_H_fuciformis_V_amabilis}), the wings still appear transparent with a bluish hue. Fluorescent microscopy observations indicate that both the transparent membrane and the veins of the wings fluoresce (Fig.~\ref{fig:Micrograph_1PF}). The latter are clearly brighter in the case of the two investigated cicada species (Fig.~\ref{fig:Micrograph_1PF}e,f), indicating that they might contain many more fluorescent molecules than the membrane. Chitin is known to give rise to autofluorescence, with a very low quantum yield~\cite{Hai2018}. In addition, autofluorescent proteins such as resilin are known to be at the origin of blue hue fluorescence emission from damselfly and dragonfly species~\cite{Gorb1999,Appel2011,Appel2015,Guillermo-Ferreira2014}. Hence both chitin and resilin are likely to play a role in the fluorescence emission observed from investigated cicada and moth species (Fig.~\ref{fig:Micrograph_1PF}e-g).
It was previously reported that the extinction coefficient $\kappa$ -namely, the imaginary part of the complex refractive index- of the wing material of \textit{C. orni} exhibits a double peak in the UV range 200-300~nm, reaching ca. $\kappa=2.5\times10^{-2}$ and $\kappa=1.5\times10^{-2}$, with a non-negligible component in the range 300-400~nm~\cite{Dellieu2014}. Similarly, Azofeifa \textit{et al.} measured peaks at 280~nm and 325~nm in the extinction coefficient for chitin samples from the exoskeleton of fresh Pacific white shrimps \textit{Litopenaeus} (formerly \textit{Penaeus}) \textit{vannamei} (Boone, 1931)~\cite{Azofeifa2012}. Our results show that light absorption in the wings of insects such as cicadas within the range 300-400~nm gives rise to fluorescence emission.

\begin{figure}
	\includegraphics[width=\columnwidth]{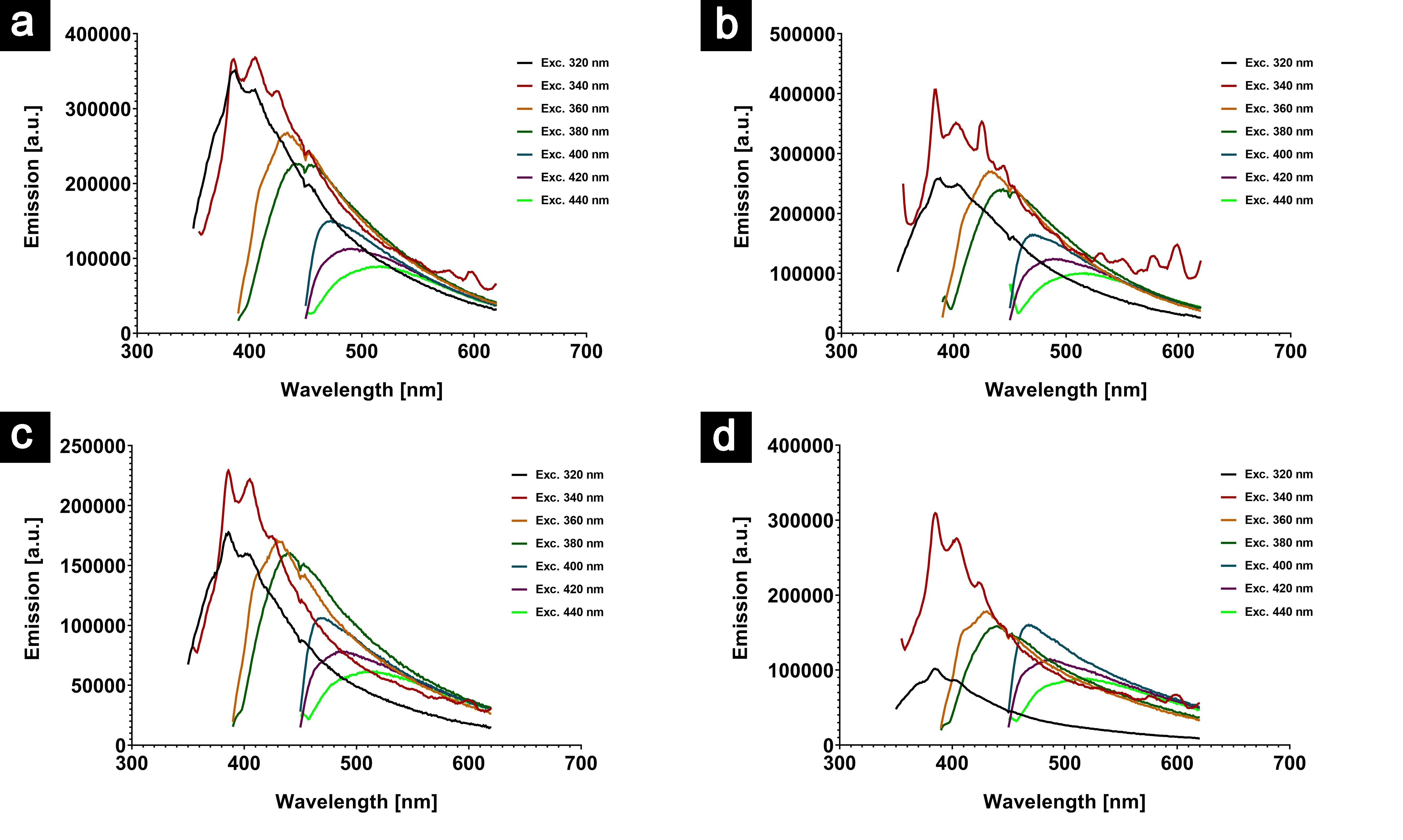}
	\caption{\textbf{Fluorescence emission from the transparent wings of a) \textit{C. orni}, b) \textit{L. plebejus}, c) \textit{H. fuciformis}, d) \textit{V. amabilis}}. The emission peak widths, intensities and positions strongly depend on the excitation wavelength and are in agreement with the narrow emission band of resilin reported in the literature~\cite{Appel2015,ANDERSEN1963}.} 
	\label{fig:Micro_1PF}
\end{figure}

In all four investigated insect wings, fluorescence emission single peaks were measured upon excitation with wavelengths ranging from 320~nm to 440~nm (Fig.~\ref{fig:Micrograph_1PF}). A 340-nm excitation wavelength appears to give rise to the most intense emission. For instance, the measured emission spectra display peaks at about 410-420~nm with a 360-nm excitation wavelength. The presence of a single peak in the emission spectra suggests that there is likely only a single type of fluorophore within the insect wings. The presence of mixtures of several fluorophores is less likely as this would imply emission spectra with multiple peaks. The measured emission peaks are similar to the fluorescence emission measured by Chuang \textit{et al.} from the wings of the damselfly \textit{Ischnura senegalensis} (Rambur, 1842)~\cite{Chuang2016}. Despite the fact that our observations are significantly different from the fluorescence response from \textit{M. pudica} male neotropical damselfly (i.e., emission at ca. 650~nm upon excitation at 405~nm), the wings of which are known to contain resilin, the emission peaks we measured are in agreement with the narrow emission band of resilin located at ca. 415~nm~\cite{Appel2015,ANDERSEN1963}.

In addition, the emission peak widths and positions strongly depend on the excitation wavelength (Fig.~\ref{fig:Micro_1PF}). The peak position shifts from ca. 380~nm to ca. 490~nm as the excitation wavelength is increased from 320~nm to 440~nm. This dependence is similar to the previously reported one for males of the beetle \textit{Hoplia coerulea} (Drury, 1773)~\cite{Mouchet2019}. It suggests that the fluorophores exhibit complex dynamics with several excited states. In complex organic fluorophores, several excited states are likely related to exciton resonances (i.e., energy transfers between different donor and acceptor groups within the same molecule)~\cite{Lakowicz1999,Ramanan2014,Mouchet2019}.

\begin{figure}
	\includegraphics[width=\columnwidth]{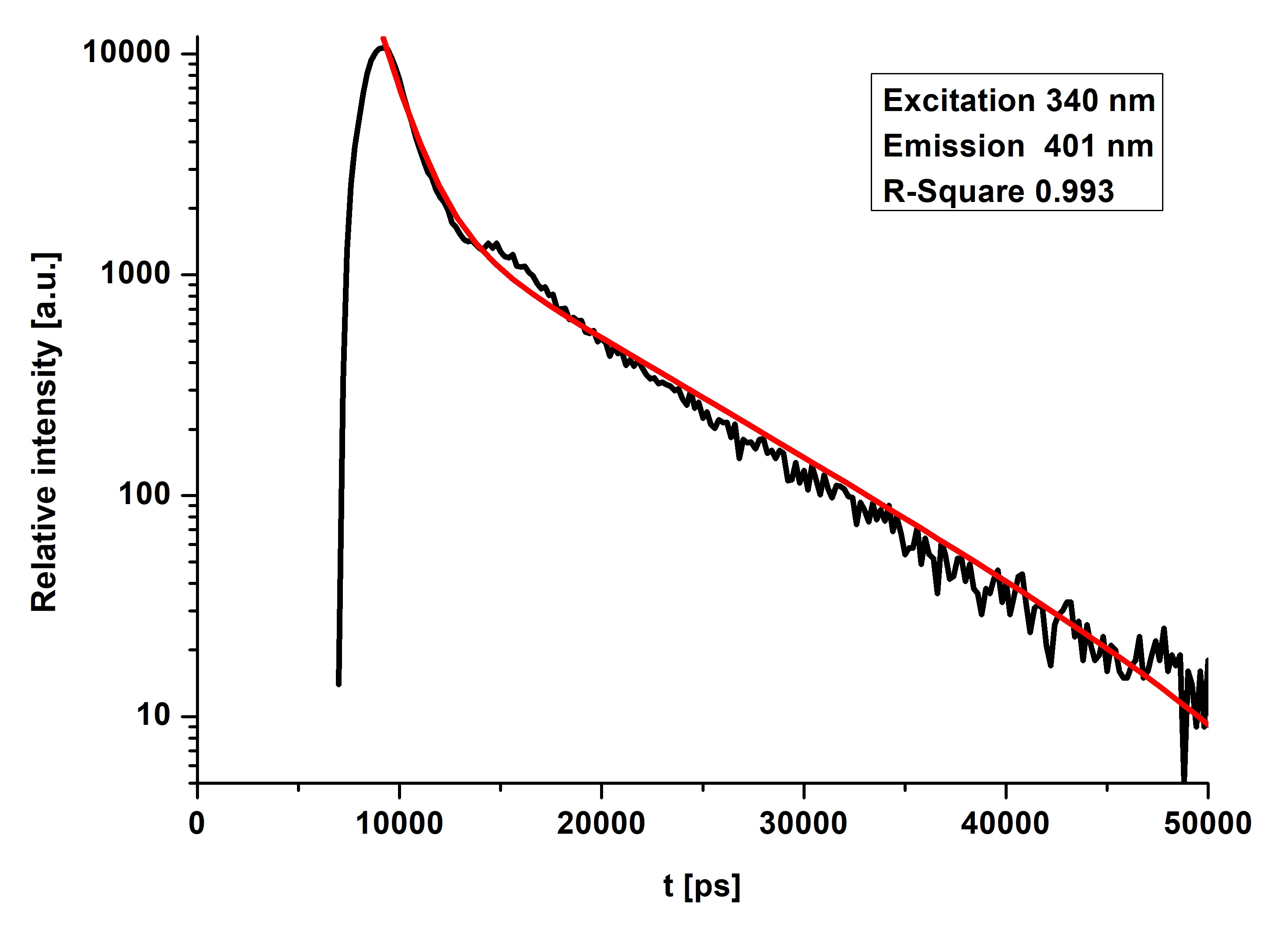}
	\caption{\textbf{Time-resolved measurement of the emission intensity of the fluorophores embedded in the wings of \textit{H. fuciformis}.} The measurements were performed with an excitation wavelength at 340~nm and an emission wavelength at 401~nm. The best fit to the decay is a double exponential function.} \label{Decay}
	
\end{figure}

In order to understand better the radiative response in the linear regime, we performed time-resolved measurement of the emitted fluorescence intensity (Fig.~\ref{Decay}).

\begin {table}
\centering

\begin{tabular}{|l|c|c|c|}
    \hline 
    \textbf{Species} & \textbf{Excitation} & \textbf{Emission}  &  \textbf{Decay time} \\ 
    & \textbf{Wavelength [nm]} & \textbf{Wavelength [nm]}  &  \textbf{[ps]} \\
    \hline 
	\textit{Cicada orni} & 340 & 385 & ${t}_{1}$ = 1271.23 \\ 
	& &  & ${t}_{2}$ = 8933.32 \\ 
	& 340 & 403 & ${t}_{1}$ = 1309.57 \\ 
	& &  & ${t}_{2}$ = 9272.42 \\ 
	& 360 & 388 & ${t}_{1}$ = 1192.96 \\ 
	& &  & ${t}_{2}$ = 8508.55 \\ 
	& 360 & 406 & ${t}_{1}$ = 1237.31 \\ 
	& &  & ${t}_{2}$ = 8831.80 \\ 
	\textit{Lyristes (Tibicen) plebejus} & 340 & 384 & ${t}_{1}$ = 1372.48 \\ 
	& &  & ${t}_{2}$ = 8288.30 \\ 
	& 340 & 402 & ${t}_{1}$ = 1184.63 \\ 
	& &  & ${t}_{2}$ = 7688.22 \\ 
	& 360 & 387 & ${t}_{1}$ = 1107.87 \\ 
	& &  & ${t}_{2}$ = 7708.26 \\ 
	& 360 & 406 & ${t}_{1}$ = 1328.55 \\ 
	& &  & ${t}_{2}$ = 8299.55 \\ 
	\textit{Hemaris fuciformis} & 340 & 384 & ${t}_{1}$ = 1322.55 \\ 
	& &  & ${t}_{2}$ = 8249.07 \\ 
	& 340 & 401 & ${t}_{1}$ = 1278.86 \\ 
	& &  & ${t}_{2}$ = 8149.58 \\ 
	& 340 & 425 & ${t}_{1}$ = 1313.99 \\ 
	& &  & ${t}_{2}$ = 8491.21 \\ 
	& 360 & 406 & ${t}_{1}$ = 1298.20 \\ 
	& &  & ${t}_{2}$ = 7997.02 \\ 
	& 360 & 424 & ${t}_{1}$ = 1284.24 \\ 
	& &  & ${t}_{2}$ = 8262.70 \\ 
	& 360 & 451 & ${t}_{1}$ = 1245.23 \\ 
	& &  & ${t}_{2}$ = 8126.60 \\ 
	\textit{Vestalis amabilis} & 340 & 384 & ${t}_{1}$ = 1235.50 \\ 
	& &  & ${t}_{2}$ = 7578.67 \\ 
	& 340 & 402 & ${t}_{1}$ = 1374.48 \\ 
	& &  & ${t}_{2}$ = 8060.57 \\ 
	& 340 & 425 & ${t}_{1}$ = 1338.49 \\ 
	& &  & ${t}_{2}$ = 8220.30 \\ 
	& 360 & 387 & ${t}_{1}$ = 1326.02 \\ 
	& &  & ${t}_{2}$ = 8226.46 \\ 
	& 360 & 406 & ${t}_{1}$ = 1203.75 \\ 
	& &  & ${t}_{2}$ = 7596.25 \\ 
	& 360 & 450 & ${t}_{1}$ = 1119.60 \\ 
	& &  & ${t}_{2}$ = 7841.73 \\ 
	\hline 
	
\end{tabular} 

\caption{\textbf{Decay times of the fluorophores embedded within the transparent wings of \textit{C. orni}, \textit{L. plebejus}, \textit{H. fuciformis} and \textit{V. amabilis}.} Measurements were performed at excitation wavelengths equal to 340~nm and 360~nm, and emission wavelengths lying within the emission peaks.}
     \label{tab:table1}
\end {table}

\newpage

In all cases (Table~\ref{tab:table1}), the best fits to double exponential functions gave rise to a long decay time (of ca. 7-8~ns) and a short decay time (of ca. 1.1-1.3~ns), whatever the excitation and detection wavelengths were. This also suggests that only one type of fluorophores is present in the wings of each insect. The long decay time is related to the radiative decay of the fluorophores, whereas the short one is related to nonradiative decay~\cite{Kolaric2010}. If we assume that measured fluorescent signals originate from the same molecule (namely, resilin), the discrepancy observed in the emission spectra and decay time among the investigated samples could be attributed to differences in the chemical environment from one species to another. 



\begin{figure}
	\includegraphics[width=\columnwidth]{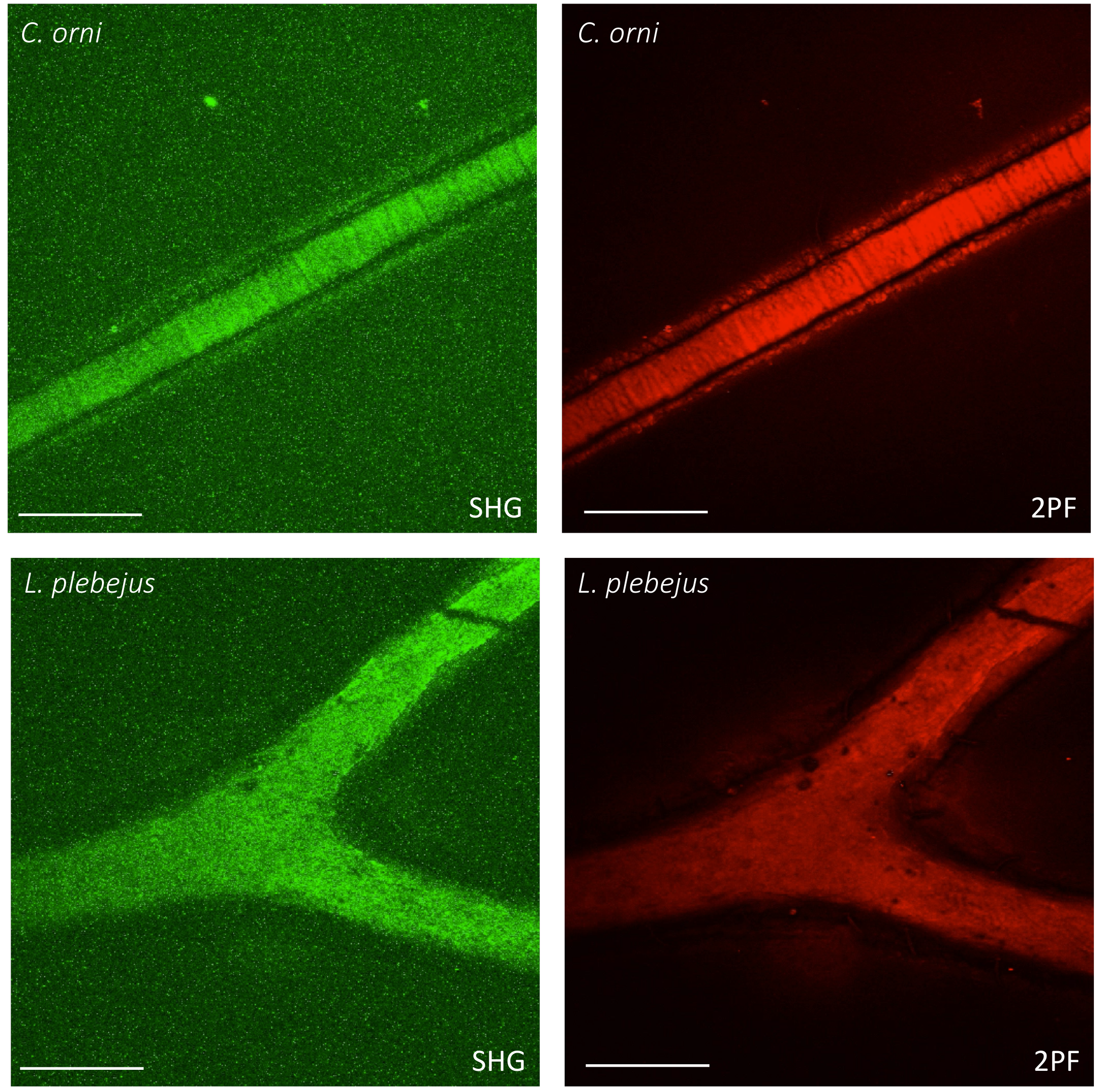}
	\caption{\textbf{Strong SHG and 2PF signals are observed from the veins of the wings.} Multiphoton micrographs of \textit{C. orni} (top) and \textit{L. plebejus} (bottom) at a fundamental wavelength of 1000~nm. Left: SHG response. Right: 2PF response. Laser power: 200~mW (\textit{C. orni}) and 80~mW (\textit{L. plebejus}). Scale bars: 200~µm.}
	\label{fig:MPMcicadas}
\end{figure}

\pagebreak

In addition, both \textit{C. orni} and \textit{L. plebejus} were probed at fundamental wavelengths ranging from 800~nm to 1100~nm, and showed significant SHG and 2PF signal response over all this wavelength range (see Fig.~\ref{fig:MPMcicadas} at a 1000-nm wavelength). At any of these selected fundamental wavelengths, the veins clearly appear in both SHG and 2PF observations. The difference in signals between the veins and the membrane (in 1PF, 2PF and SHG) is likely due to the significant difference in composition~\cite{Gorb1999,Appel2011,gullion2017}: the veins of cicadas would show a much higher concentration in chitin and resilin. The former is known to give rise to an SHG signal~\cite{rabasovic2015nonlinear}. Additionally, the veins are much thicker than the membrane, leading to an increased amount of material. Optical second-order behaviours such as SHG are known to take place exclusively in non-centrosymmetric media~\cite{verbiest2009second}. The detected SHG signal implies that the veins are materially organised in a non-centrosymmetric way. 


From a visual point of view, the biological significance of fluorescence emission in the transparent wings of insects depends on the total number of photons emitted or scattered from the wings and transmitted through the wings from their backgrounds: many cicadas are known to be mainly active during the hottest hours of summer days, when sunlight is very intense but also at twilight and during the night. Although the measured fluorescence emission ranges match the range of both scotopic and photopic photoreceptors of many insects~\cite{Peitsch1992,Stavenga2002,Stavenga2009,Lunau2014}, it is hard to conclude that the measured 1PF signal plays any role in cicadas' behaviour since the quantum yield of natural fluorophores are generally low~\cite{Iriel2010a,Iriel2010b,Taboada2017}, specifically the one of chitin~\cite{Hai2018}, and the ratio of UV light in both sun- and moonlight is very limited with respect to the one of visible light (for instance, ca. 6\% of sunlight at sea level lies in the UV range, whereas about 50\% is visible)~\cite{Moan2001}. The role of this fluorescence emission may possibly be a by-product of another biological function such as photoprotection due to the increased absorption of UV by the fluorophores (for example, when the wings are at rest over the insect's body in the case of cicadas) and incidental property of resilin that enhances wing flexibility.


\section{Conclusion}

Transparent wings of two cicada species, namely \textit{C. orni}, \textit{L. plebejus}, as well as the broad-bordered bee hawk-moth \textit{H. fuciformis}, were demonstrated to emit light under incident UV light by fluorescence decay. All of them showed similar 1PF properties, as their wings were excited in the 320-440~nm range and emitted in the range 380-490~nm. They were compared to the transparent wings of the damselfly \textit{V. amabilis}, which showed similar fluorescent behaviours. Dameselflies fluoresecence emission is reported to arise from substances such as resilin embedded in their tissues. Additionally, the wings of both cicada species and more specifically their veins showed strong SHG and 2PF signals, which are possibly due to the higher concentration of resilin or chitin in the veins with respect to the membrane that makes up most of the wing. The nonlinear optical characteristics of the cicadas' wings gave us more insight into the wing structure, indicating that multiphoton microscopy add valuable information while characterising insect integuments. 




\textbf{Acknowledgments}
SRM was supported by the Belgian National Fund for Scientific Research (FRS-FNRS) (91400/1.B.309.18F), the Maturation Fund of the Walloon Region, and a BEWARE Fellowship of the Walloon Region (Convention n°2110034), as a Postdoctoral Researcher. DM acknowledges KU Leuven Postdoctoral Mandate Internal Funds (PDM) for a Postdoctoral felowship (PDM/20/092). TV acknowledges financial support from the Hercules Foundation. BK acknowledges financial support from the "Action de Recherche Concert\'{e}e" (BIOSTRUCT project‚ No.10/15-033) of UNamur, from Nanoscale Quantum Optics COST-MP1403 action and from FRS-FNRS; Interuniversity Attraction Pole: Photonics@be (P7-35, Belgian Science Policy Office).

\bibliographystyle{unsrt} 
\bibliography{references}

\end{document}